\documentclass[aps,prb,twocolumn,groupedaddress,floatfix]{revtex4-2}

\usepackage{graphicx}
\usepackage{dcolumn}
\usepackage{bm}

\begin{document}

\title{Error analysis of bit-flip qubits  under  random telegraph noise for low and high temperature measurement application}
\author{Win Wang, Sanjay Prabhakar${^1}$}
\affiliation{
Department of Natural Science, Gordon State College, 419 College Drive, Barnesville, GA 30204
}
\date{Feb 09, 2020}

\begin{abstract}
Achieving small error for qubit gate operations under random telegraph noise (RTN) is of great interest for potential applications in quantum computing and quantum error correction. I numerically estimate the error generated in the qubit driven by $\pi$, CORPSE, SCORPSE, symmetric and asymmetric pulses in presence of RTN. For a special case when pulse acts in x-direction and RTN in z-direction, I find that for small value of noise correlation time, $\pi$-pulse has small error than all the other pulses. For large value of noise correlation time, symmetric pulse generates small error  for small energy amplitudes of noise strength, whereas CORPSE pulse has small error  for large energy amplitudes of noise strength. For the pulses acting in all the three directions, several pulse sequences were identified that generate small error in presence of small and large strength of energy amplitudes of RTN. More precisely, when $\pi$ pulse acts in x direction, CORPSE pulse acts in y direction and SCORPSE pulse acts in z-direction induce small error and may consider for better candidate in implementing  of bit-flip quantum error correction. Error analysis of small energy amplitudes of RTN may be useful for low temperature measurements,  whereas  error analysis of large energy amplitudes of RTN may be useful for room temperature measurements of quantum error correction codes.
\end{abstract}



\maketitle


\section{Introduction}
Qubits can be manipulated in a desired fashion by excellent architect design in several physical devices, such as, quantum dots, cavity quantum electrodynamics, superconducting devices, Majorana fermions~\cite{choi00,burkard99,hu00,koh12,xiao10,prabhakar10,prabhakar14,prabhakar14a,
vool16,lau17,yoshihara17,clarke16}. Manipulation of qubits in these devices seems promising in that one can make  quantum logic gates and memory devices for various quantum information processing applications. Such devices require sufficiently short gate operation time combined with long coherent time~\cite{loss98,golovach04,amasha08,balocchi11}. When a qubit is operated on by a classical bit, then its decay time is given by a relaxation time which is also supposed to be longer than the minimum time required to execute one quantum gate operation.

In most cases, compared to coherent time, the dephasing time of qubits in presence of noise is reduced by several orders of magnitude due to coupling of qubits to the environment. The reduction of dephasing time depends on the specific dynamical coupling sequence from where  the principle of quantum mechanics is inevitably lost.   Therefore, one might need to decouple the qubits from the environment and may consider more robust topological  method to preserve a quantum state against noise, enabling robust quantum memory\cite{june15}.
Hence, to make quantum computers, one needs to find an efficient and experimentally feasible algorithm that overcome  the issues  of undesired interactions of qubits to the surrounding environment. Interactions of qubits to the surrounding environment destroy the quantum coherence that lead to generate errors and loss of fidelity. In quantum computing language,  this phenomenon is called decoherence. For example, experimental observations reported that in GaAs quantum dots, decoherence time, $T_2^\star \approx 10 ns$ and coherent time, $T_1 \approx 0.1 ms$, whereas  for Si, $T_2^\star \approx 100 ns$ and $T_1 \approx 0.1 ms$~\cite{bluhm11,petta05,maune12,pla12,malinowski18,friesen17,martins17,he19,geng16,liu19,throckmorton17,yang18neural,huang17}. There are several possible ways to overcome the issues of decoherence, as for example, fidelity recovery by applying error-correcting codes, decoherence free subspace coding, noiseless subsystem coding, dynamical decoupling from hot bath, numerical design of pulse sequences, that is  more robust to experimental
inhomogeneities, and optimal control pulse based on Markovian
master equation descriptions.~\cite{ofek16,michael16,brown16,reiserer16,lau16,liu16,castro12,wang12}
In Ref.~\onlinecite{mottonen06}, authors provided a scheme of high fidelity recovery of qubits operating by $\pi$, CORPSE, and  SCORPSE pulses in random telegraph noise pulses    but present work is different in that qubits operating by symmetric pulse shows the smallest error generation in the regime of small energy amplitudes of  noisy strength that has large noise correlation time. In addition, we investigate the interplay of these pulses acting in all three (x,y,z) directions while RTN still acts in z-direction and then identified those pulses which provide less error on bit-flip qubits. Identifying pulse sequences that generate small error in small energy amplitudes of noise strength may be suitable for the experiments operating at low temperatures  whereas identifying pulse sequences that generate small error in large energy amplitudes of noise strength may be suitable for the experiments operating at high  temperatures.

In this paper, we consider the design of several control pulses acting on
a single bit-flip computational basis states in a noisy environment. The present work seek to identify  different regimes of operating parameters in the designed control pulses   that eliminate the series of phase and dynamical errors and induce the recovery of  high fidelities. The calculations are restricted to only eliminate the phase errors, which are more robust due to stochastic time-varying amplitudes, appear  in the model Hamiltonian. More precisely, the designed pulses are $\pi$, CORSPE, SCORPSE, symmetric and asymmetric acting on a qubit in a  random telegraph noise environment. Then checking the quantum gate errors at various noise correlation times as well as various energy amplitudes of noise strength as an indication of most efficient way to perform algorithm for quantum bit operations. The results of calculations shows that when the qubits are driven by pulses in the x-direction and the noises act in the z-direction then the symmetric pulse sequence along with small  energy amplitude of noise strength, (i.e., $\Delta \approx 0.125 \hbar/a_{max}$)  induce less systematic errors than all the other pulses ($\pi$, CORPSE, SCORPSE and asymmetric pulses). Such error analysis in the qubits is useful for the laboratory experiments operating at low temperatures where one can correct the systematic errors in a more efficient way by designing additional quantum gates in a physical device . On the other hand for the case of strong noise environments (i.e., energy amplitude of noise strength, $\Delta \approx 0.25 \hbar/a_{max}$), may suitable for the experiments operating at room temperature, CORPSE  pulse is the most efficient way to reduce the error.
For a more general case, I consider the pulses acting in arbitrary in  x, y and z directions and show that when $\pi$ pulse acts in x direction, CORPSE pulse acts in y direction and SCORPSE pulse acts in z-direction in presence of arbitrary low and high temperature measurements noise condition have large fidelity recovery and may consider for implementing in future for electronic circuits design to minimize error.

The paper is organized as follows. In section~\ref{theoretical-model}, we provide a theoretical description of the model Hamiltonian for a qubit operating under several control pulses in a random telegraph noise environment. In section~\ref{results-discussions}, we analyze two main results: (i) qubits driven by a pulse in the x-direction and RTN noise acts in z-direction. (ii) individual pulse acts in the x,y and z-direction in the qubits and RTN still acts in the z-direction. Finally I conclude our results.

\section{Model Hamiltonian}\label{theoretical-model}
\begin{figure*}
\includegraphics[width=18cm,height=5cm]{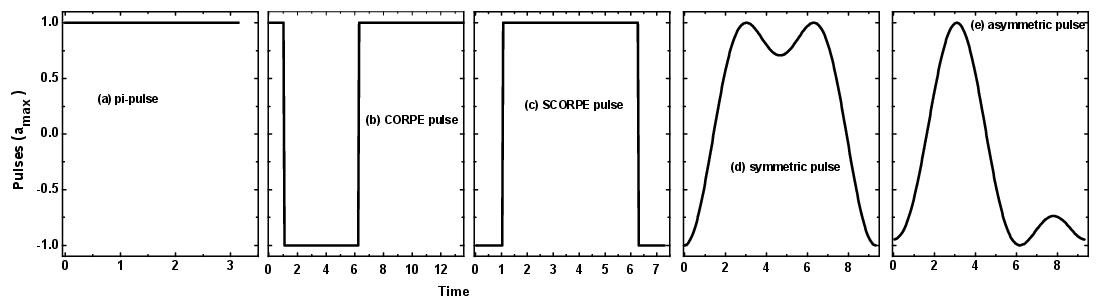}
\caption{\label{fig1} One qubit is driven by (a) $\pi$, (b) Corspe, (c) Scorspe, (d) symmetric and (e) asymmetric pulses. These pulses are under consideration to achieve fidelity recovery (or, minimize the error) in one qubit gate operation under random telegraph noise.
}
\end{figure*}

We write the effective Hamiltonian of a single qubit as~\cite{nelson}
\begin{equation}
H(t)=\sum_{i\in \{x,y,z\}} \frac{1}{2}\left[a_i(t)+ \eta_i(t)\right]\cdot \sigma_i,
\label{Ht}
\end{equation}
where $a_i(t)$ are the energy amplitude of the external control fields,  $\eta_i(t)$ are the energy amplitude of the random environmental noise and $\sigma_i$ is the Pauli spin matrices. We further assumed that the designed control pulses are acting on all three x,y and z-directions while environmental noises are acting only in z-direction. Hence, the effective Hamiltonian~(\ref{Ht}) can be written as
\begin{equation}
H(t)= \frac{1}{2}\left[a_x(t)\sigma_x + a_y(t)\sigma_y + a_z(t)\sigma_z + \eta_z(t)\sigma_z\right].
\label{Hxz}
\end{equation}
Several designed pulses are shown in Fig.~\ref{fig1}, which have the following sequences for energy amplitudes variations:

For $\pi$ pulse,
\begin{equation}
a_\pi(\theta)=a_{0},~~\mathrm{for}~~ \theta \in \left[0,\pi\right].
\label{a-pi}
\end{equation}
For CORPPE pulse,
\begin{eqnarray}
a_C(\theta)  &=& a_{max},~~\mathrm{for}~~ 0 < \theta < \frac{\pi}{3},\nonumber\\
        &=& -a_{max},~~\mathrm{for}~~ \frac{\pi}{3} \leq  \theta \leq 2\pi,\nonumber\\
        &=& a_{max},~~\mathrm{for}~~ 2\pi <  \theta \leq \frac{13\pi}{3}.
\label{a-c}
\end{eqnarray}
For SCORPSE pulse,
\begin{eqnarray}
a_{SC}(\theta)  &=& -a_{max},~~\mathrm{for}~~ 0 < \theta < \frac{\pi}{3},\nonumber\\
        &=& a_{max},~~\mathrm{for}~~ \frac{\pi}{3} \leq  \theta \leq 2\pi,\nonumber\\
        &=& -a_{max},~~\mathrm{for}~~ 2\pi <  \theta \leq \frac{7\pi}{3}.
\label{a-sc}
\end{eqnarray}
For symmetric pulse,
\begin{eqnarray}
a(t) = \frac{\pi}{2}+\left(a-\frac{\pi}{2}\right) \cos{\left(\frac{2\pi t}{\tau}\right)} -a \cos{\left(\frac{4\pi t}{\tau}\right)},\label{at}\\
a_{sym}(t)=\frac{a_{max}}{N_1} a(t)-a_{max},
\label{a-sym}
\end{eqnarray}
where  $a_{max}$=1, $a$ =  5.263022(1/$\tau$), $\tau$=9.325  and 1/$N_1$=0.8477 are  integers.

For asymmetric pulse,
\begin{widetext}
\begin{equation}
a_{asym}(t)=\frac{a_{max}}{N_2} \left[a(t) +  b \sin{\left(\frac{2\pi t}{\tau}\right)} - \frac{b}{2} \sin{\left(\frac{4\pi t}{\tau}\right)}\right]-a_{max},
\label{a-sym}
\end{equation}
\end{widetext}
where  $b$ = 17.850535(1/$\tau$) and 1/$N_2$=0.412 are  integers. Above pulse sequences, which are  plotted in Fig.~\ref{fig1},  are used to drag the qubits in noisy environments to seek the recovery of the measurement of high fidelity quantum gates.

\begin{figure*}
\includegraphics[width=15cm,height=8cm]{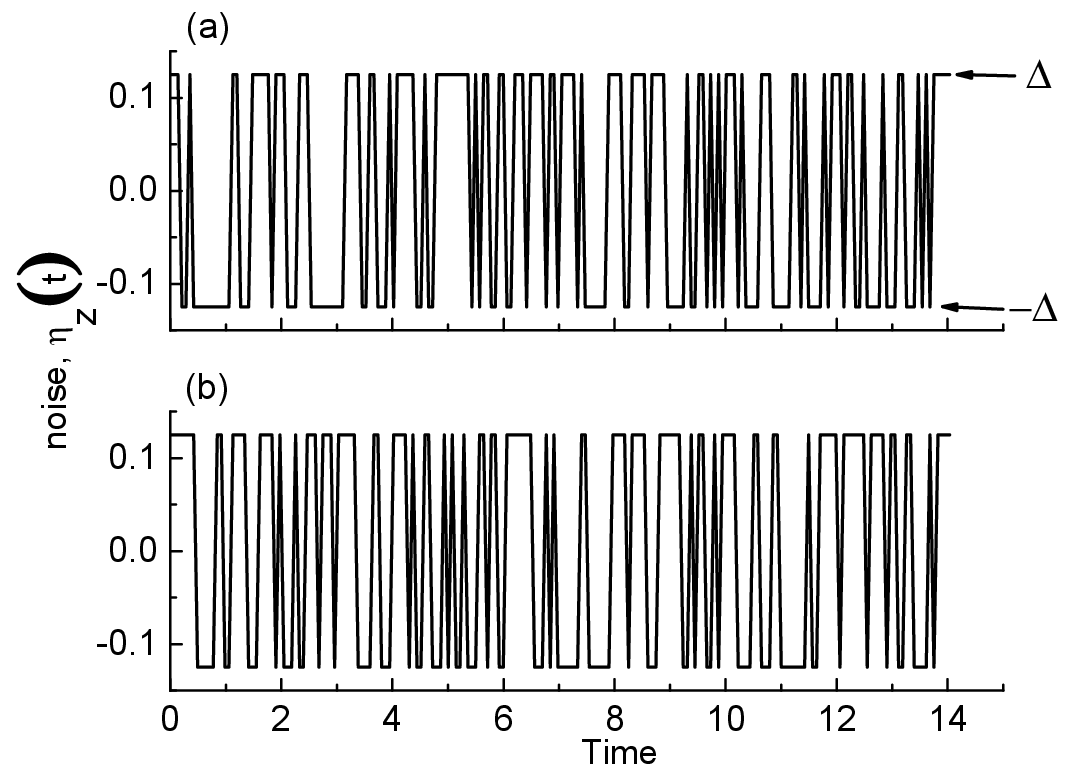}
\caption{\label{fig2} Simulations of random telegraph noise (RTN) as a function of time ($\hbar/a_{max}$) for the correlation time, $\tau_c$= 0.001($\hbar/a_{max}$).  Note that the density of RTN jumps between $\pm \Delta$ is random which is shown in Fig.\ref{fig2}(a) and (b). Here only two noises functions are shown for demonstration purpose but in realistic simulations of finding error, 600 RTN trajectories have been chosen for realistic simulations of finding fidelity (see Fig.~\ref{fig3})
}
\end{figure*}

\begin{figure}
\includegraphics[width=8.5cm,height=7cm]{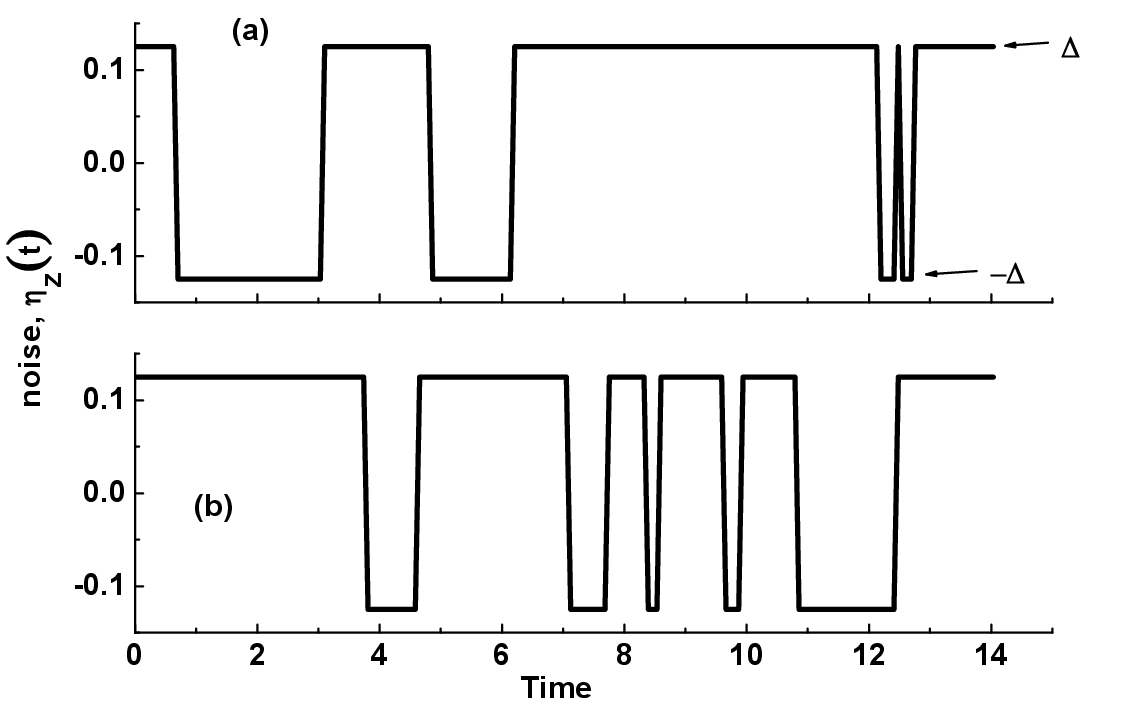}
\caption{\label{fig3} Same as to Fig.~\ref{fig2} but $\tau_c$=1($\hbar/a_{max}$).
}
\end{figure}

For random environmental noise, $\eta_z(t)$ changes randomly between two values $\eta(0)=\Delta$ and $\eta(0)=-\Delta$, where $\Delta$ is the energy amplitude of the strength of the noise. The environmental noise   function is written as
\begin{equation}
\eta_z(t)= (-1)^{\Sigma_i \Theta\left(t-t_i\right)} \eta(0),
\label{eta-t}
\end{equation}
where $\Theta(t)$ is the heaviside step function, and $\tau_c$ is the correlation time. The jump time instants $t_i$ is expressed as
\begin{equation}
t_i= \sum_{j=1}^i -\tau_c ln(p_j),
\label{ti}
\end{equation}
where the random numbers $p_i\in (0,1)$ determine the sample trajectories of random telegraph noises. Two randomly generated environmental noise functions are shown in Fig.~\ref{fig2} and \ref{fig3}. For the simulations of fidelity measurement, $500$ randomly generated noise functions are chosen. As can be seen in Fig.~\ref{fig2}, there are large density of random jump noises in the vicinity of zero correlation times, $\tau_c$. On the other hand, as $\tau_c$ increases, the density of random jump noises decreases that can be seen in Fig.~\ref{fig3}.

Suppose the effective Hamiltonian~(\ref{Hxz}) is acting only on the qubit. Hence, to find the system dynamics,  average over different noise trajectories were chosen to find the system dynamics. Therefore, the density matrix of the dynamics of the system is written as
\begin{equation}
\rho(t)=lim ~N \rightarrow \infty~ \frac{1}{N} \sum_{k=1}^N U_k \rho_0 U_k^\dagger,
\label{rho}
\end{equation}
where $\rho_0$ is the initial state of the system and $\{U_k \}$ is the unitary time evolution of the qubit under the influence of control pulses (see Fig.~\ref{fig1}) and randomly generated noise functions (see Fig.~\ref{fig2a} and \ref{fig2b}). The  unitary time evolution operator is written as
\begin{equation}
U_k=\tau e^{{-i/2\hbar}\int_0^t d\tau \sum_{i\in \{x,y,z\}} \left[a_i(t) \cdot \sigma_i + \eta_z(t)\sigma_z\right] },
\label{Uk}
\end{equation}
where $\tau$ is the time ordering parameter.  Considering $\rho_f$ is the final state of the system then the fidelity for the system is given by
\begin{equation}
\phi= tr\{\rho_f^\dagger \rho_T \},
\label{phi}
\end{equation}
where $\rho_T$ is the final desired state of the qubit.

\begin{figure*}
\includegraphics[width=18cm,height=7cm]{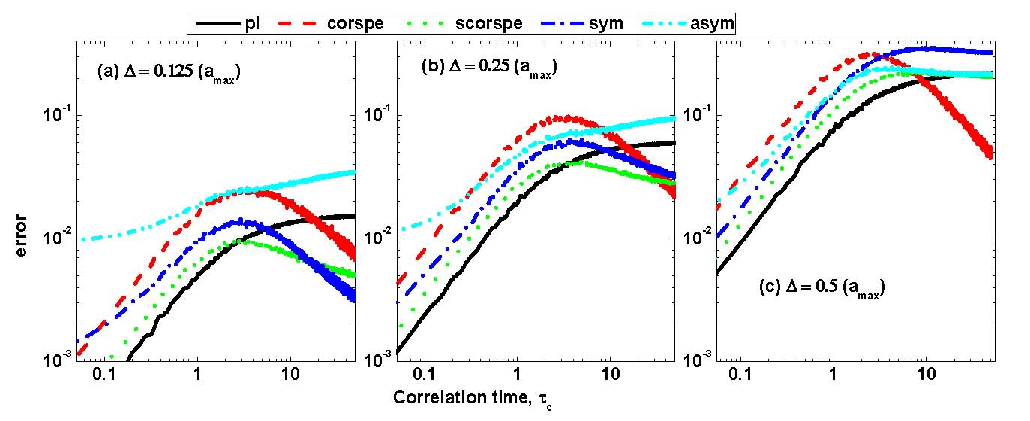}
\caption{\label{fig4} error (1-$\phi (\rho_f,\rho_0)$) vs noise correlation time, $\tau_c (\hbar/a_{max})$  for $\pi$ pulse (black, solid line), corspe pulse (red, dashed line), scorspe pulse (green, dotted line) and symmetric pulse (blue, dotted line). We chose $\Delta=0.125a_{max}$ as the amplitude of strength of the RTN noise.
}
\end{figure*}
\begin{figure*}
\includegraphics[width=18cm,height=6cm]{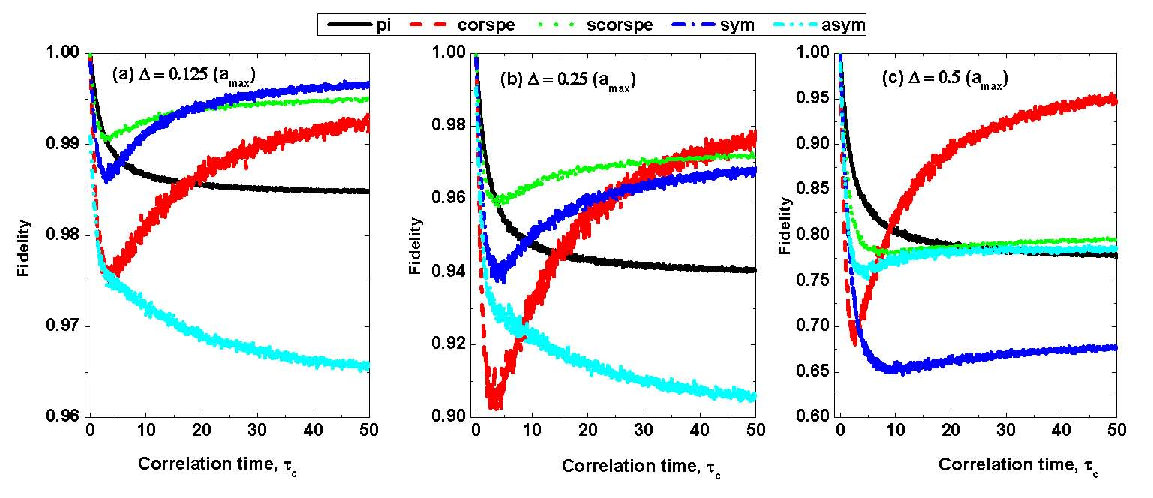}
\caption{\label{fig5}  Bit flip fidelities, $\phi (\rho_f,\rho_0)$ vs noise correlation time, $\tau_c (\hbar/a_{max})$  for $\pi$ pulse (black, solid line), corspe pulse (red, dashed line), scorspe pulse (green, dotted line), symmetric pulse (blue, dashed-dotted line) and asymmetric pulse (magenta, dashed-dotted-dotted line). Notice that the recovery of fidelity is the largest for symmetric pulse in Fig.\ref{fig4}(a) (i.e., $\Delta = 0.125$) but corpe pulse in Fig.\ref{fig4}(b) (i.e., $\Delta = 0.25$) in the vicinity of $\tau_c \rightarrow \infty$.
}
\end{figure*}
\begin{figure*}
\includegraphics[width=18cm,height=8cm]{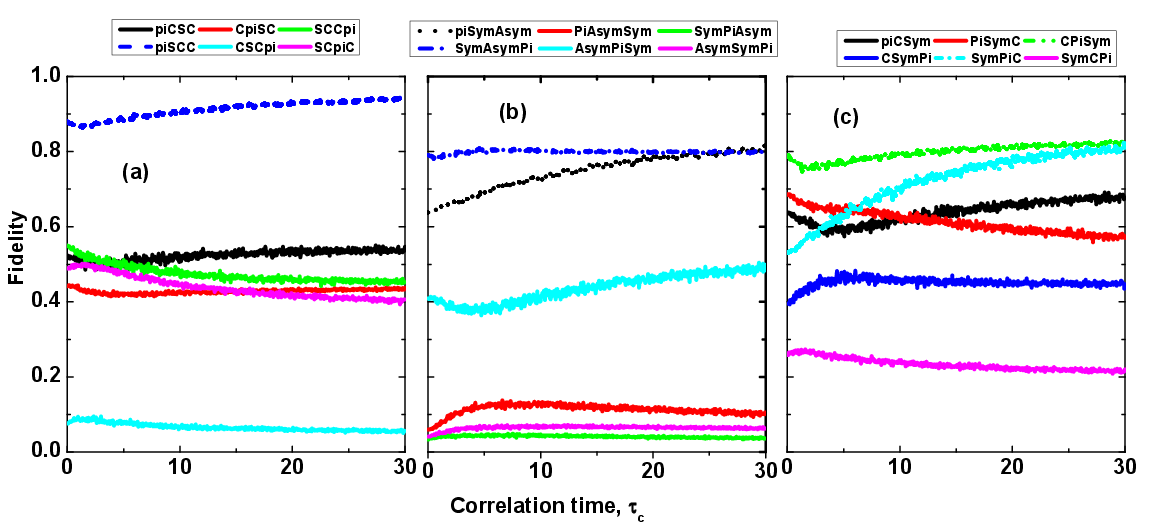}
\caption{\label{fig6}  Bit flip fidelities, $\phi (\rho_f,\rho_0)$ vs noise correlation time, $\tau_c (\hbar/a_{max})$  for possible combinations of pulse sequences $\{a_x(t),a_y(t),a_z(t)\}$ in the Hamiltonian~\ref{Hxz}. Notice that the recovery of fidelity is the largest for $(a_\pi(t),a_{SC},a_C)$ (Fig.~\ref{fig6}(a), dashed line, blue), $(a_\pi(t),a_{sym},a_{Asym})$ (Fig.~\ref{fig6}(b), dotted line, black) and  $(a_{sym}(t),a_{Asym},a_{\pi})$ (Fig.~\ref{fig6}(b),dashed-dotted line, blue). Similarly, the recovery of fidelity is the largest for $(a_C(t),a_\pi,a_{sym})$ (Fig.~\ref{fig6}(c). All other combinations of pulse sequences have small fidelities that may be useless for performing experiments. Here we chose  $\Delta = 0.125$.
}
\end{figure*}

\begin{figure*}
\includegraphics[width=18cm,height=8cm]{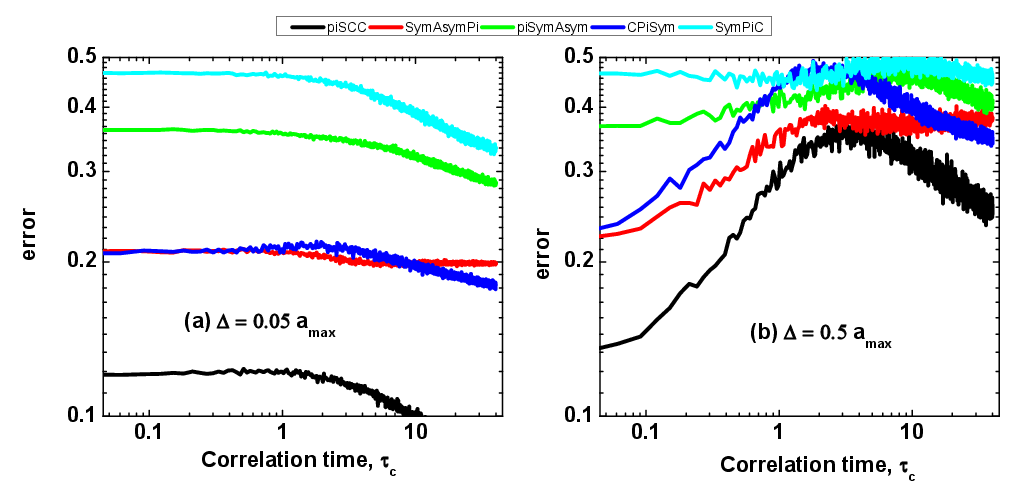}
\caption{\label{fig7}  Bit flip fidelities, $\phi (\rho_f,\rho_0)$ vs noise correlation time, $\tau_c (\hbar/a_{max})$  for $\Delta = 0.05$ (left panel) for low temperature measurements and $\Delta= 0.5$ (right panel) for large temperature measurements. The combinations of pulse sequences $\{a_x(t),a_y(t),a_z(t)\}$ in the Hamiltonian~\ref{Hxz} $\pi$ are  chosen same as to Fig.~\ref{fig6} but only consider those pulse sequences which has larger fidelity recovery again noises.
}
\end{figure*}

\section{Computational Methods}\label{cm}

We consider the bit flip computational basis states $|0>$ and $|1>$ as:
\begin{equation}
\rho_0=\left(\begin{array}{cc} 0 & 0\\ \nonumber
0 & 1 \end{array}\right),~~~\rho_f=\left(\begin{array}{cc} 1 & 0\\ \nonumber
0 & 0 \end{array}\right).
\end{equation}

For the limiting case as the correlation time, $\tau_c\rightarrow 0$ and $\Delta \rightarrow \infty$,
$\Delta^2\tau_c/2 \equiv \hbar^2 \Gamma$
remains finite value where $\Gamma$ is the decoherence time. In this case RTN model reduces to white noise. One may apply  Markovian master equation in the model Hamiltonian to find the fidelity of the system~\cite{blum}. In this paper,  I do not apply Markovian master equation but write a python code that simulate the RTN trajectories numerically. I have chosen $500$ RTN trajectories and express the density matrix of the system by a unitary quantum trajectory approach that is valid for all values of the correlation times $\tau_c$. Throughout the paper, I have chosen $\hbar = a_{max} = 1$.

\section{Results and Discussions}\label{results-discussions}

Considering $a_y(t) = a_z(t) = 0$ in \ref{Hxz} and utilizing Eq.~\ref{phi}, the influence  of $\pi$, CORSPE, SCORPSE, symmetric and assymetric pulses on the error correction and fidelity measurement of qubit under random telegraph noises is shown in Figs.~(\ref{fig4}) and (\ref{fig5}). Here $\delta = 0.125 a_{max}$, $\delta = 0.25 a_{max}$  and $\delta = 0.5 a_{max}$  are chosen(a),(b) and (c) of Figs.~(\ref{fig4}) and (\ref{fig5}), respectively.  In the regime of vanishing noise correlation time ($i.e., \tau_c \rightarrow 0$), error is minimum  shown  in Fig.~\ref{fig4}(a,b,c) or Fidelity is maximum shown in Fig.~\ref{fig5})(a,b,c) for $\pi$ pulse followed by SCORPE, CORPE, symmetric and asymmetric pulses. The small error or large fidelity for $pi$ pulse in the qubit operation  is due to the fact that the single qubit under $\pi-$pulse  does not have enough time to drift into the direction of  densely populated random noise (see Fig.~\ref{fig2}). Note that the noise function is very dense in the vicinity of zero correlation time (see Fig.\ref{fig2}) while noise function has no jumps in the vicinity of infinite correlation time (e.g., density of noise jumps decreases as $\tau_c$ increases from \ref{fig2} to \ref{fig3}). The energy amplitude of the noise functions rapidly changes its sign between  $\pm \Delta$. Hence, in the vicinity of zero correlation time, the error is smaller or the fidelity is larger to be about 99$\%$ for $\pi-$pulse than all the other pulses (e.g.,  scorpe, corpe, symmetric and asymmetric).

As a correlation time increases, the density of noises or the frequency of random jumps of noise functions decreases that can be seen in Figs.~(\ref{fig2}) and (\ref{fig3}). Here I chose $\tau_c = 0.001 \hbar/a_{max}$ in Fig.\ref{fig2} and ($\tau_c = 1 \hbar/a_{max}$) in Fig.~(\ref{fig3}). From Figs.~\ref{fig2} and Fig.\ref{fig3}, one can conclude that in the regime of sufficiently large noise correlation time ($i.e., \tau_c \rightarrow \infty$), the frequency of random jumps of noise function $\eta(t)$ is significantly reduced, where noise function can be treated as a constant (no jumps). The gate pulses acting on this regime with $\Delta = 0.125$ quickly recover their lost fidelities except for $\pi-$ and asymmetric pulses (see Fig.~\ref{fig3}(a) and \ref{fig4}(a)). However, for large value of $\Delta = 0.5$, only CORPE pulse has higher fidelity recovery, which can be utilized to correct systematic time-independent errors.
Since gate pulses (except $\pi$ pulse) change its amplitude between $\pm a_{max}$ with the variation of gate operation time, one can find the  global minimum point at  $\tau_c \approx 3\hbar/a_{max}$, from where the lost fidelity starts to recover.

I have plotted the error vs noise correlation time with $\Delta = 0.25$ in Fig.~\ref{fig4}(b) and with $\Delta = 0.5$ in Fig.~\ref{fig4}(c)  or fidelity vs noise correlation time with $\Delta=0.25{a_max}$ in \ref{fig5}(b) and with $\Delta=0.5{a_max}$ in \ref{fig5}(c). Since $\Delta=$0.25 and 0.5 may consider  a strong noisy environments, the qubit operation in this case may be suitable for gate operation at room temperature, the CORSPE pulse sequence recovers its lost fidelities more rapidly than other pulse sequences due to its  large gate operation time over all the other pulses.  Hence in a very noisy environment, CORSPE pulse is only suitable candidate that correct the systematic errors. For a less noisy environment, i.e.,  $\Delta = 0.125$ in Fig.~\ref{fig4}(a) and \ref{fig5}(a), which may suitable for gate operation at low temperature, $\pi-$pulse in the regime of $\tau_c \rightarrow 0$ and  the symmetric pulse sequences in the regime of $\tau_c \rightarrow \infty$ are the most suitable pulse to recover their lost fidelities.

Now I consider the pulses acts in x, y and z-directions while the RTN noises still act in the z-direction. The description of the qubits in this situations are well formulated by the Hamiltonian~(\ref{Hxz}). I have tested interplay of several possible combinations of the  pulse sequences $\{a_x(t), a_y(t), a_z(t)\}$ among $(a_\pi(t)$, $a_C(t)$ and $a_{SC}(t)$, $a_{sym}(t)$ and $a_{asym}(t)$ in the Hamiltonian~(\ref{Hxz}) and then plotted  the fidelity vs correlation time in Fig~\ref{fig6}. The results from this plots show that the pulses $a_x(t)$ represented by $a_\pi(t)$, $a_y(t)$ represented by $a_C(t)$ and $a_z(t)$ represented by $a_{SC}(t)$ has large value of fidelity shown in Fig.~\ref{fig6}(a). Similarly, the pulses $a_x(t)$ represented by $a_\pi(t)$, $a_y(t)$ represented by $a_{sym}(t)$ and $a_z(t)$ represented by $a_{Asym}(t)$ has large value of fidelity shown in Fig.~\ref{fig6}(b)(dotted-line). In a similar way,  the pulses $a_x(t)$ represented by $a_{sym}(t)$, $a_y(t)$ represented by $a_{Asym}(t)$ and $a_z(t)$ represented by $a_\pi(t)$ has large value of fidelity shown in Fig.~\ref{fig6}(b)(dotted-line). All the other combinations of pulse sequences has lower fidelity and may be useless for achieving less error in the  low and high temperature measurements of bit-flip qubits.

In Fig.~\ref{fig7}, I consider the same pulses that has large fidelity measurement in Fig.~\ref{fig6} but with different values of noise strength. Note that small value of noise strength, $\Delta$ is applicable for low temperature measurements while large value of $\Delta$ is applicable for high temperature measurements, probably room temperature.
Here I find that when $\pi$ pulse acts in x direction, CORPSE pulse acts in y direction and SCORPSE pulse acts in z-direction in presence of arbitrary low and high temperature measurements noise condition have large fidelity recovery and may consider for implementing in future for electronic circuits design to minimize error.

\section{Conclusion}

In Figs.~(\ref{fig3}) and (\ref{fig4}) we have demonstrated a possible way to minimize error and achieve high fidelity quantum gate operations using several bounded control pulses (e.g., $\pi$, CORPSE, SCORPE, symmetric and asymmetric pulses) acting on a one qubit system in presence of random telegraph noises. Here we have compared the fidelities obtained from the pulses discussed above. We conclude that in the limit of vanishing noise correlation time, $\pi$-pulse can be used for the measurement of achieving high fidelity due to its small gate operation time. In the limit of large correlation time, two pulses namely symmetric pulse and CORPSE pulse were identified to achieve recovery of high fidelity. More precisely, symmetric pulse (see Fig.~\ref{fig3}(a)) provide large fidelity for the small energy amplitudes of noise strength which may be useful for the experiments that perform at low temperatures. On the other hand, CORPE pulse (see Fig.~\ref{fig3}(b,c)) provide large fidelity for the large energy amplitudes of noise strength which may be useful for the experiments that perform at high temperatures.
Finally in Fig.~\ref{fig7}, I have shown  that when $\pi$ pulse acts in x direction, CORPSE pulse acts in y direction and SCORPSE pulse acts in z-direction in presence of arbitrary low and high temperature measurements noise conditions, large fidelity recovery can be achieved and may consider for implementing in future for electronic circuits design to minimize error.


%

\end{document}